\documentstyle[epsfig,12pt]{article}
\def\beqn{\begin{eqnarray}}
\def\eeqn{\end{eqnarray}}
\def\beq{\begin{equation}}
\def\eeq{\end{equation}}

\def\cR{{\cal R}}

\def\calN{{\cal N}}
\def\calO{{\cal O}}
\def\calP{{\cal P}}

\def\xbj{x_{\rm Bj}}
\def\xp{x_{\cal P}}

\begin{document}
\pagestyle{empty}
\begin{flushright}
{ETH-TH/96-27}\\
{DTP/96/74}\\
July 1996 \\
\end{flushright}
\vspace*{5mm}
\begin{center}
{\large {\bf Hard Diffractive Scattering: Partons and QCD$^*$}}\\
\vspace*{1cm}

{\bf Z.~Kunszt} \\
\vspace{0.3cm}
Theoretical Physics, ETH--H\"onggerberg,\\
 Zurich, Switzerland \\
\vspace{0.5cm}
{\bf W.J.~Stirling} \\
\vspace{0.3cm}
Departments of Physics and Mathematical Sciences,  University of Durham, \\ 
Durham, United Kingdom \\
\vspace{2cm}
{\bf ABSTRACT} \\ \end{center}
\vspace*{3mm}
\noindent
The HERA diffractive structure function data
are interpreted in terms of `diffractive parton distributions' which 
satisfy DGLAP evolution. These distributions are modeled assuming
that the scattering takes place off a colour singlet `pomeron'
target. A quantitative test of the universality of diffractive parton
distributions is proposed.
\vfill
\begin{flushleft}
$^*$ Based on talks presented at the Workshop on HERA Physics,
Durham, September 1995 and  the Workshop on Deep Inelastic Scattering and 
Related Phenomena (DIS96), Rome, April 1996.
\end{flushleft}
\eject
\setcounter{page}{1}
\pagestyle{plain}

Measurements of deep inelastic scattering events with a large rapidity
gap at HERA~\cite{gaporig,H1lastyear,ZEUSlastyear,H1thisyear,ZEUSthisyearMX} 
have generated renewed interest in the
idea of `diffractive hard scattering'. It now appears that in
a significant fraction of these events there is a diffractively
scattered  proton in the final state~\cite{ZEUSthisyearLPS}. One can therefore
introduce the idea of `diffractive parton distributions', i.e.
 parton distributions in the hadron under the condition that
the hadron is diffractively scattered. Just as the
total deep inelastic structure function, measured in the
process $\gamma^* p \to X$, can be written as
a sum over parton distributions times a short-distance part,
$
F_2(\xbj,Q^2) = \sum_q\; \int dx\; f_{q/p}(x,\mu^2) \; \widehat{F}_{2q}
( \xbj /  x, Q^2, \mu^2 )$ with
$\widehat{F}_{2q}  = e_q^2\; \delta (1- \xbj/ x )
+ \calO(\alpha_s) $,
we may define a similar decomposition for the diffractive structure
function, measured in the
process $\gamma^* p \to p  X$,
\begin{equation}
{dF_2^D(\xbj,Q^2;\xp,t)\over d\xp d t} = \sum_q\; \int dx\;
{d f_{q/p}(x,\mu^2; \xp , t) \over d \xp d t} \;  \widehat{F}_{2q}
\left( {\xbj\over x}, Q^2, \mu^2 \right) .
\label{eq:f2d}
\end{equation}
Here $1-\xp$ ($\xp \ll 1$) and $t$ ($ \vert t \vert \ll Q^2$)
 are respectively the longitudinal energy fraction
and the $t$--channel momentum transfer of the proton in the final state.
The quantities  $d f_{q/p}(x,\mu^2; \xp , t)  / d \xp d t $  can then
be regarded as diffractive parton distributions.\footnote{Similar
quantities (`fracture functions') were introduced in
Ref.~\cite{TRENVEN}.}
A rather detailed theoretical discussion of such diffractive structure
functions has recently been presented in Ref.~\cite{SOPER}. In
particular it has been shown that an operator definition exists,
and that the diffractive distributions should satisfy the same (DGLAP)
evolution equations as the usual parton distributions.
However it is  not yet clear  whether the concept of short-distance
factorization generalizes to {\it any} diffractive hard scattering
process, for example the production of
large-mass Drell-Yan lepton pairs or large $E_T$ jets 
in hadron-hadron collisions~\cite{CFS}. Even if factorization is violated for
diffractive hard scattering in such collisions, the effect may be  weak 
at high energy. It therefore seems to us not unreasonable  to assume the 
approximate validity  of universal factorization and to test its
consequences.
The purpose of the present study is to explore further the idea
of hard diffractive factorization by using the HERA $F_2^D$
 data to model diffractive parton distributions, and
then to use these to make quantitative predictions for other
diffractive hard-scattering processes, in particular for the production
of $W$ bosons in $p \bar p$ collisions at the 
Tevatron.

An important property of the HERA diffractive events~\cite{gaporig}
 is the approximate
{\it factorization} of the structure function $F_2^D$ (integrated over $t$)
 into a function of $\xp$ times a function of
  $\beta = \xbj/\xp$:
$F_2^{D} \sim \xp^{-n} {\cal F}(\beta,Q^2)$.
This, together with the observed rapidity-gap topology
of the events,  strongly suggests that the deep inelastic scattering takes
place off a slow-moving colourless target $\calP$ `emitted'
by the proton, $p \to \calP p$, and with a fraction $\xp \ll 1$
of its momentum.\footnote{This physical picture refers to the infinite
momentum frame of the proton. Although  it is rather different, it is
not {\it a priori} inconsistent with the so-called aligned
quark picture~\cite{beejay} which is valid in the rest frame of the proton.}
 If this emission is described by a universal flux function
$f_{\calP}(\xp, t) d\xp dt $, then the diffractive structure function $F_2^D$
is simply a product of this and the structure function of the
colourless object, $F_2^\calP(\beta,Q^2)$. Since the scattering evidently
takes place off point-like  charged objects, we may write the latter as a sum
over quark-parton distributions,
i.e. $ F_2^{\calP}(\beta,Q^2) = \beta \sum_q e_q^2 q_{\calP}(\beta,Q^2)$,
 in leading order.
In this way we obtain a model for the diffractive parton distributions
introduced above:
\begin{equation}
{d f_{q/p}(x,\mu^2; \xp , t) \over d \xp d t}
 =     f_{\calP}(\xp, t)\;   f_{q/\calP}(\beta=x/\xp,\mu^2)\; .
\label{eq:fqp}
\end{equation}

Many recent studies have analysed the HERA diffractive structure
function data using this theoretical framework. A popular
choice is to assume that the colour-neutral target is the Regge
pomeron, in which case the emission factor $f_\calP$ is already
 known from soft hadronic physics (for a review see Ref.~\cite{PVL}).
In its simplest version, this model would predict
$
 F_2^{D}(x,Q^2; \xp,t) = f_{\calP}(\xp,t)\;F_2^{\calP}(\beta,Q^2)
$
with $f_{\calP}(\xp,t) = F_{\calP}(t) \xp^{2\alpha_{\calP}(t)-1}$,
i.e. a factorized structure function with $n \approx  2\alpha_{\calP}(0)-1
\approx 1.16$. This model is based on the notion of `parton constituents
in the pomeron' first proposed by Ingelman and Schlein~\cite{IS} and supported
by data from UA8~\cite{UA8}. 
In such a model, a modest amount of factorization breaking,
such as that observed in the more recent H1 and ZEUS analyses 
\cite{H1thisyear,ZEUSthisyearMX,ZEUSthisyearLPS},
could be accommodated by  invoking a sum over Regge trajectories, each
with a different intercept and structure function:
\beq
 F_2^{D}(\beta,Q^2; \xp,t) = \sum_{\cR}\;
 F_{\cR}(t) \xp^{2\alpha_{\cR}(t)-1}\;F_2^{\cR}(\beta,Q^2)\; ,
\label{eq:reggeons}
\eeq
which would yield an effective $n$ which depends on $\beta$
but is approximately independent of $Q^2$.\footnote{For a recent
quantitative study see Ref.~\protect\cite{GBK}.}
 Note that since in practice
the variables $\xp$ and $t$ are integrated over, what is measured is
a linear combination of $F_2^\cR$ structure functions or, equivalently,
the parton distributions in an effective colour-neutral target:
\beq
\int d \xp dt \, F_2^{D}(\beta,Q^2; \xp,t) = \sum_{\cR}\;
A_\cR F_2^{\cR}(\beta,Q^2) = \beta \sum_q e_q^2\; \sum_{\cR}\;
A_\cR  q_{\cR}(\beta,Q^2) ,
\eeq
where the coefficients $A_\cR$ are independent of $\beta$ and $Q^2$.
Since the DGLAP equations are {\it linear} in the parton distributions,
the $Q^2$ evolution of the integrated  structure function $F_2^D$ 
should also be calculable perturbatively.\footnote{In principle, for 
the integrated structure function there is a correction to the evolution
equations from the small but finite probability that the final-state
proton is produced in the hard-scattering process. However if the integration
is only over a limited region in $t$, as in the present
context, this correction will be completely
negligible.} 
In the present study we assume, for simplicity,
 pomeron exchange only (i.e. $\cR = \calP$)  and use
 the parametrization of Ref.~\cite{DL} for $ \alpha(t) $ and $F_\calP(t)$.
The $\xp$ dependence of the diffractive structure
function  predicted by this type of `soft pomeron' model
is roughly consistent within errors
with the H1~\cite{H1thisyear} and ZEUS~\cite{ZEUSthisyearMX,ZEUSthisyearLPS}
 data, although there is some indication from the latter that a somewhat
steeper $\xp$ dependence is preferred.

Various models for the parton distributions in the
`pomeron'   $f_{q/\calP}(x,\mu^2)$  have been proposed,
ranging from the two extremes of mainly gluons to mainly quarks.
Recent studies in the framework of perturbative QCD
DGLAP evolution can be found in Refs.~\cite{H1thisyear,CKMT,GS,JPP,GK,GP}.
As we shall demonstrate below, models of both types can
be  constructed to agree with the HERA data. A key issue
concerns the existence of a momentum sum rule for the colourless
object $\calP$. This is a matter of some dispute, and there is indeed
no theoretical proof that such a sum rule should exist. Of course
because it is the {\it product} of $f_{q/\calP}$ and $f_\calP$
that appears in the expression for the structure function, one can
simply impose a momentum sum rule on the parton distributions and
absorb an overall normalization $\calN$, unchanged by $Q^2$ evolution,
 into $f_\calP$. This is the
approach we shall adopt here.

Since our purpose is to explore the consequences of a parton
interpretation of deep inelastic diffractive scattering for other
processes, we consider three qualitatively different models
of diffractive parton distributions. The first is purely quark-like
at the chosen starting scale $\mu_0^2$, with gluons generated only
at higher scales through standard DGLAP evolution. The second
contains a mixture of quarks and gluons at $\mu_0^2$, and the third
is predominantly gluonic. Since in each case the quark content
is constrained by structure function data, and since we choose
to impose a momentum sum rule, the overall normalization factors
$\calN$ are different in the three models (see below).
In each case the starting distributions are chosen to give
satisfactory agreement with the H1~\cite{H1lastyear}
 and ZEUS~\cite{ZEUSlastyear} data. Since the errors on these data
 are  quite large, it is not necessary to perform detailed
multiparameter fits. The starting distributions (at $\mu_0^2 =
2 $~GeV$^2$) are given in the following Table.

\begin{center}
\begin{tabular}{|c|c|c|c|}  \hline
\rule[-1.2ex]{0mm}{4ex} Model & $q(x,\mu_0^2)$ & $g(x,\mu_0^2)$ & $\calN$ \\ 
\hline
\rule[-1.2ex]{0mm}{4ex} 1   & $ 0.314 x^{1/3}(1-x)^{1/3} $  &  0  &  1.62    \\
\rule[-1.2ex]{0mm}{4ex} 2   & $ 0.2 x(1-x)$   &  $4.8 x(1-x)$   &  2.85    \\
\rule[-1.2ex]{0mm}{4ex} 3   & $ 0.081x(1-x)^{0.5}$  &  $ 9.66x^8(1-x)^{0.2}$ 
 &  1.57    \\
\hline
\end{tabular}
\end{center}

\noindent Here $q$ refers to an individual light-quark distribution with
SU(3) flavour symmetry assumed, i.e. $q = u = \bar u = d = \bar d
= s = \bar s$, so that  the momentum sum rule constraint at $\mu_0^2$
is $\int_0^1 dx\; x(6q+g)=1$. Charm quarks are generated by massless
DGLAP evolution ($g \to c \bar c$) at higher scales.
The Table also lists the normalization
factors required to fit the data. We emphasize that these are correlated
to our use of the Donnachie-Landshoff parametrization of $f_\calP(\xp,
t)$.\footnote{We integrate the $\xp$
and $t$ variables over the range appropriate for the HERA
experiments.}  Different emission factors will lead to different normalizations.
\begin{figure}[htb]
\begin{center}
\hspace{-5.5truecm}
\vspace{1.0truecm}
\mbox{\epsfig{figure=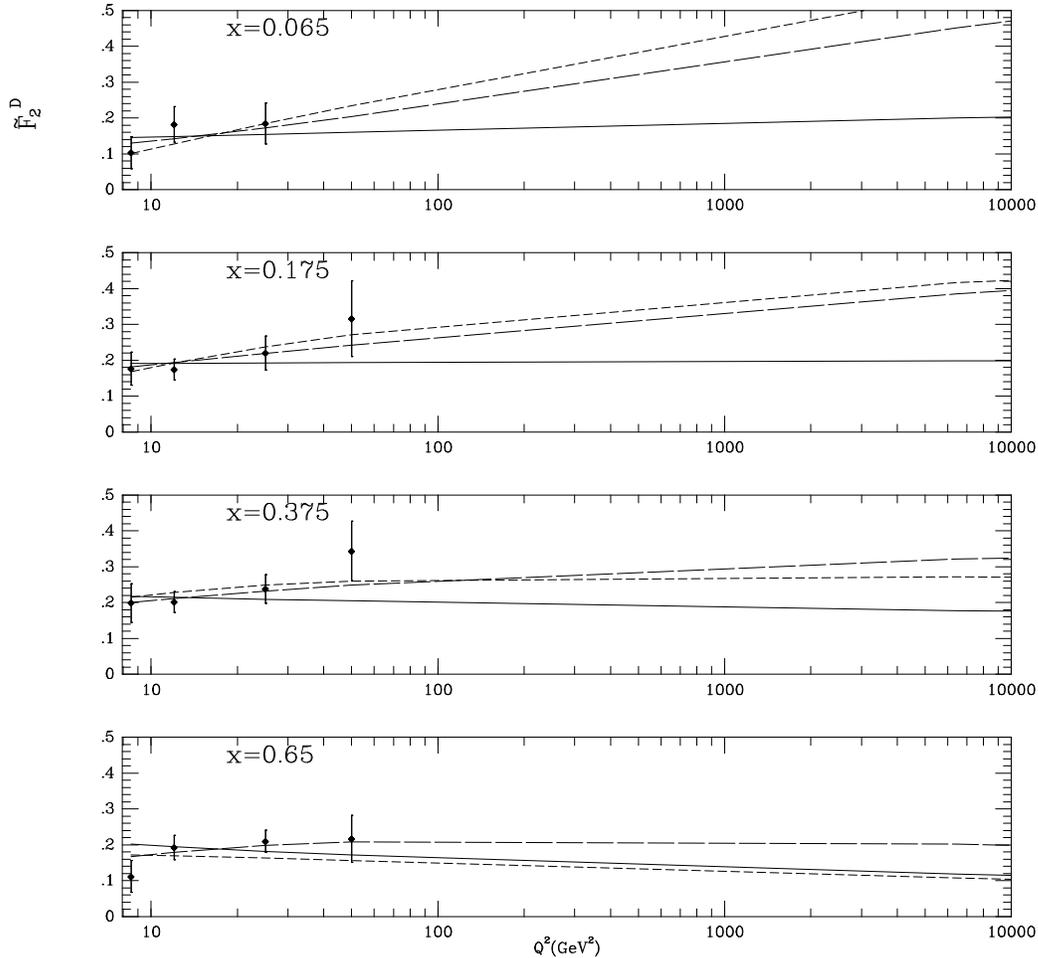,height=10cm,angle=270}}
\vspace{-5truecm}
\caption{Fits (solid line: Model~1, dashed line: Model~2,
long-dashed line: Model~3) to the H1 $F_2^D$ data~\protect\cite{H1lastyear}.}
\label{fig:rpomfig}
\end{center}
\end{figure}
Fig.~\ref{fig:rpomfig} shows the quality of the fits to the H1 diffractive
structure function data \cite{H1lastyear}. The fits to the ZEUS data
are comparable in quality and are not shown. More details about the fits
and the different $Q^2$ evolution in the three models will be 
given elsewhere~\cite{POMKS}.

The concept of `universal pomeron structure' can be tested in hard diffractive
processes in hadron-hadron collisions. The cleanest process to study
would appear to be weak boson production at the Tevatron $p \bar p$
collider. This was first
studied  in Ref.~\cite{BRUNI}, where it was estimated
that the single diffractive component of the total $W$ cross section
could be as large as $20\%$. More recently, the CDF
collaboration~\cite{GOULIANOS} has searched for such diffractive events,
and derived a preliminary upper limit on the single
diffractive cross section of  `a few per cent'.
Using the three models discussed above, it is straightforward to
compute the single diffractive $W$ cross section:
\begin{equation}
\sigma^{SD}(W) \sim 2 {\bar q}_{\bar p} \otimes \calN f_\calP q_{\calP} ,
\end{equation}
where the factor of two corresponds to either the proton or antiproton
being quasi-elastically scattered. To avoid uncertainties from
high-order corrections it is sensible to normalize this prediction to
the total $W$ cross section, $\sigma^{tot}(W) \sim {\bar q}_{\bar p} \otimes
q_p$. For the parton distributions in the proton we use the MRS(A$'$) 
set~\cite{MRSG}. The factorization scale is chosen to be $Q=M_W$, and we sum
over (four) flavours of quarks in the initial-state proton and pomeron.
Following Ref.~\cite{BRUNI}, we define `single diffractive' events
by $x_\calP < 0.1$, integrating over all $t$. In practice, 
the events are defined by rapidity gaps of a certain minimum size,
and therefore the observed diffractive cross section must be corrected
to the theoretical prediction based on, say, $x_\calP < 0.1$ using
a Monte Carlo simulation~\cite{GOULIANOS}.
With the above choice of parameters and cuts we find
\begin{equation}
{\sigma^{SD}(W) \over \sigma^{tot}(W)} = \left\{
\begin{array}{ll}
5.3\% & \mbox{Model 1} \\
6.5\% & \mbox{Model 2} \\
7.4\% & \mbox{Model 3}
\end{array}   \right.
\label{SDW}
\end{equation}
The important point to note here is that the model predictions are quite
similar: the hypothesis of hard diffractive factorization has yielded
a well-constrained prediction for the single diffractive cross section.
This result can be understood in terms of the evolution of the
quark densities in the pomeron. The single diffractive $W$ cross
section at the Tevatron samples the quarks in the pomeron at
$\langle x_{q/\calP} \rangle \sim 0.4$. At low $Q^2$ the quark
distributions at this $x$ value
are constrained by the HERA $F_2^D$ data to be roughly the same. As
$Q^2$ increases the  distributions diverge, reflecting the
quantitatively different
gluon contributions to the DGLAP evolution. However at $Q^2 \sim
10^{4}$~GeV$^2$, the relevant value for $W$ production, the difference
between the quarks in the three models is still not very large, 
and the predictions for $\sigma^{SD}(W)$ are correspondingly similar, see
Fig.~1.\footnote{Using the formalism developed above, 
predictions can also be made for diffractive heavy 
quark~\cite{HEYSSLER} and Higgs~\cite{HIGGS,POMKS} production.}
We stress that failure to observe a diffractive $W$ cross section
of the order of the values given in Eq.~(\ref{SDW}) would 
cast  serious doubt on the `universal pomeron structure' hypothesis.

In summary, we have shown how the HERA diffractive structure function data
can be understood in terms of diffractive parton distributions, which
satisfy DGLAP evolution and can be modelled in terms of various
 combinations of quarks and gluons in an effective colour-neutral
target. We have discussed the concept of the universality
of such distributions, and shown how the measurement of the single
diffractive $W^\pm$ cross section at the Tevatron will provide a stringent
test of the universality property.

\section*{Acknowledgements}
 This work was
supported in part by the EU Programme
``Human Capital and Mobility'', Network ``Physics at High Energy
Colliders'', contract CHRX-CT93-0357 (DG 12 COMA).


\begin{thebibliography}{99}
\bibitem{gaporig}
ZEUS collaboration: M.~Derrick et al., {\em Phys. Lett. } {\bf B315} (1993) 481;
{\bf B332} (1994) 228; {\bf B338} (1994) 483. \\
H1 collaboration: T.~Ahmed et al., {\em Nucl. Phys. } {\bf B429} (1994) 477.
\bibitem{H1lastyear} H1 collaboration: T.~Ahmed et al.,  {\em Phys. Lett. }
{\bf B348} (1995) 681.
\bibitem{ZEUSlastyear} ZEUS collaboration: M.~Derrick et al.,  {\em Z. Phys. }
{\bf C68} (1995) 569.
\bibitem{H1thisyear} H1 collaboration: P.R.~Newman, 
presented at the  Workshop on Deep Inelastic Scattering and 
Related Phenomena (DIS96), Rome, April 1996.
\bibitem{ZEUSthisyearMX}ZEUS collaboration: H.~Kowalski,
presented at the  Workshop on Deep Inelastic Scattering and 
Related Phenomena (DIS96), Rome, April 1996.
\bibitem{ZEUSthisyearLPS}ZEUS collaboration: E.~Barberis, 
presented at the  Workshop on Deep Inelastic Scattering and 
Related Phenomena (DIS96), Rome, April 1996.

\bibitem{TRENVEN} L.~Trentadue and G.~Veneziano, {\it Phys. Lett.}
{\bf B323} (1994) 201.
\bibitem{SOPER} A.~Berera and D.E.~Soper, {\it Phys. Rev.} {\bf D53} (1996)
6162.\\
A.~Berera, preprint PSU/TH/169 (1996).

\bibitem{CFS}J.C.~Collins, L.~Frankfurt and M.~Strikman, 
{\it Phys. Lett.} {\bf B307} (1993) 161. \\
 A.~Berera and D.E.~Soper, {\it Phys. Rev.} {\bf D50} (1994)
4328.
\bibitem{beejay}J.D.~Bjorken, Proc. Int. Workshop on Deep Inelastic Scattering 
  and Related Subjects, Eilat, Israel (1994). \\
W.~Buchm\"uller and A.~Hebecker, {\it Phys. Lett.} {\bf B355} (1995) 573.

\bibitem{PVL} P.V.~Landshoff, Cambridge  preprint DAMTP 96/48 (1996).

\bibitem{IS}
G.~Ingelman and P.~Schlein,  Phys. Lett. {\bf B152} (1985) 256.

\bibitem{UA8}
UA8 collaboration: R.~Bonino et al., {\it Phys. Lett.} {\bf B211} (1988) 239;
A.~Brandt et al., {\it Phys. Lett.}  {\bf B297} (1992) 417.

\bibitem{GBK}K.~Golec-Biernat and J.~Kwiecinski, INP Cracow preprint
1734/PH (1996).

\bibitem{DL}
A.~Donnachie and P.V.~Landshoff, {\it Phys. Lett.} {\bf B191} (1987) 309;
{\bf B198} (1987) 590(E).

\bibitem{CKMT}
A.~Capella, A.~Kaidalov, C.~Merino and J.~Tran~Thanh~Van,
{\it Phys. Lett.} {\bf B343} (1995) 403; A.~Capella et al., 
{\it Phys. Rev.} {\bf D53} (1996) 2309. 
\bibitem{GS}
T.~Gehrmann and W.J.~Stirling, {\em Z. Phys. } {\bf C70} (1996) 89. 
\bibitem{JPP}
J.P.~Phillips,  Proc. Workshop on Deep Inelastic Scattering and QCD,
Paris, France (1995), eds. J.F.~Laporte and Y.~Sirois,  p.~359;
 preprint DESY-95-152C. 
\bibitem{GK}
K.~Golec-Biernat and J.~Kwiecinski, {\it Phys. Lett.} {\bf B353} (1995) 329.
\bibitem{GP}
K.~Golec-Biernat and J.P.~Phillips, {\em J. Phys. G} {\bf 22} (1996) 921.

\bibitem{POMKS}Z.~Kunszt and W.J.~Stirling, in preparation.

\bibitem{BRUNI}
P.~Bruni and G.~Ingelman, {\it Phys. Lett.} {\bf B311} (1993) 317.

\bibitem{GOULIANOS}
CDF collaboration: presented by K.~Goulianos
at the 10th Topical Workshop on Proton-Antiproton Collider
Physics, Batavia IL, May 1995, preprint
FERMILAB-CONF-95-244-E (1995).

\bibitem{MRSG}
A.D.~Martin, R.G.~Roberts and W.J.~Stirling, 
{\it Phys. Lett.} {\bf B354}  (1995) 155.

\bibitem{HEYSSLER}M.~Heyssler, Durham preprint DTP/96/10 (1996).

\bibitem{HIGGS} 
A.~Sch\"afer, O.~Nachtmann and R.~Sch\"opf,
{\it Phys. Lett.} {\bf B249} (1990) 331. \\
 D.~Graudenz and G.~Veneziano, {\it Phys. Lett.} {\bf B365} (1996) 302.



\end{thebibliography}
\end{document}